\documentclass[10pt,twoside]{IEEEtran}

\usepackage{etoolbox}
\newtoggle{doublecolumn}
\newtoggle{calculateFigures}

\toggletrue{doublecolumn}
\togglefalse{calculateFigures}

\usepackage{etex}

\usepackage[english]{babel}
\usepackage{lipsum}
\usepackage{amsbsy,amsmath,amsfonts,amssymb,amsthm}
\usepackage{mathtools}
\usepackage{textcomp} 
\usepackage{relsize}

\usepackage{subdepth} 

\usepackage{stackengine}

\usepackage{bm,cite}
\usepackage{cases}
\usepackage{times,url,verbatim}

\usepackage[noend]{algpseudocode}
\usepackage{booktabs}
\usepackage{tabularx,array,dcolumn,multirow}

\usepackage{enumitem} 

\usepackage{algorithm}

\usepackage{yfonts}

\usepackage{xr} 

\usepackage{soul} 

\usepackage{blkarray,bigdelim} 

\allowdisplaybreaks[4]

\usepackage{centernot} 

\newcommand{\subparagraph}{}

\iftoggle{doublecolumn}{
    \newtheorem{thm}{Theorem}
    \newtheorem{fact}{Fact}
    \newtheorem{lemma}{Lemma}
    \newtheorem{definition}{Definition}
    \newtheorem{conj}{Conjecture}
    \newtheorem{propos}{Proposition}
    \newtheorem{corol}{Corollary}
    \newtheorem{ass}{Assumption}
    \newtheorem{example}{Example}
    \newtheorem{remark}{Remark}
    \newtheorem{note}{Note}
    \newtheorem{obs}{Observation}
}{

    \newtheoremstyle{exampstyle}
      {0} 
      {0} 
      {\itshape} 
      {} 
      {\bfseries} 
      {.} 
      {.5em} 
      {} 

    \theoremstyle{exampstyle} 
    \theoremstyle{exampstyle} 
    \theoremstyle{exampstyle} 
    \theoremstyle{exampstyle} 
    \theoremstyle{exampstyle} 
    \theoremstyle{exampstyle} 
    \theoremstyle{exampstyle} 
    \theoremstyle{exampstyle} 
    \theoremstyle{exampstyle} 
    \theoremstyle{exampstyle} 
    \theoremstyle{exampstyle} 
    \theoremstyle{exampstyle} 
}

\makeatletter
\newcommand{\pushright}[1]{\ifmeasuring@#1\else\omit\hfill$\displaystyle#1$\fi\ignorespaces}
\newcommand{\pushleft}[1]{\ifmeasuring@#1\else\omit$\displaystyle#1$\hfill\fi\ignorespaces}

\begingroup
\catcode`\#=11

\endgroup

\usepackage{graphicx,xcolor,float,dblfloatfix}
\usepackage{psfrag}

\usepackage{caption}
\usepackage{subcaption}

\usepackage[compact]{titlesec} 

\newcommand{\subalign}[1]{%
  \vcenter{%
    \Let@ \restore@math@cr \default@tag
    \baselineskip\fontdimen10 \scriptfont\tw@
    \advance\baselineskip\fontdimen12 \scriptfont\tw@
    \lineskip\thr@@\fontdimen8 \scriptfont\thr@@
    \lineskiplimit\lineskip
    \ialign{\hfil$\m@th\textstyle##$&$\m@th\textstyle{}##$\crcr
      #1\crcr
    }%
  }
}

\graphicspath{%
{./Figures/}
}

\usepackage{color}

\usepackage{caption}
\usepackage{subcaption}
\begin{document}

\title{Multicast Transmissions in \\Directional mmWave Communications}

\author{Alessandro~Biason and Michele~Zorzi\\
\IEEEauthorblockA{\small Department of Information Engineering, University of Padova - via Gradenigo 6b, 35131 Padova, Italy}\\
\IEEEauthorblockA{\small email: \{biasonal,zorzi\}@dei.unipd.it}
\vspace{-.5cm}
}

\maketitle
\pagestyle{empty}
\thispagestyle{empty}


\begin{abstract}
Multicast transmissions have been widely analyzed in traditional networks as a way to improve spectrum efficiency when multiple users are interested in the same data. However, their application to mmWave communications has been studied only marginally so far. The goal of this paper is to partially fill this gap by investigating optimal and suboptimal multicast schemes for mmWave communications with directional beams. In particular, we propose a Markov setup to model the retransmission status of the unsuccessfully transmitted packets and, because of the computational complexity of the optimal solution, we introduce a suboptimal hierarchical optimization procedure, which is much easier to derive. Finally, we numerically show that restricting the link to unicast beams is strongly suboptimal, especially when many packets have to be transmitted.
\end{abstract}

\section{Introduction}

Wireless communication using millimeter waves (mmWaves) is considered a game changer paradigm, which promises to satisfy the ever growing data rate requirements of the mobile terminals~\cite{Niu2015}. With mmWaves, the communication frequencies shifted from few GHz to tens or hundreds of GHz (e.g., $30-300$~GHz). This has the twofold consequence of increasing the available bandwidth and, simultaneously, decreasing the antenna size (and thus allow the integration of large arrays of antennas in a small chip area). 
On the other hand, multicast communications have been proven to be beneficial and to increase the bandwidth efficiency in many different scenarios~\cite{Vella2013}. Nevertheless, a joint analysis of mmWave and multicast communications has been investigated only marginally so far. Our goal is to advance the state of the art and introduce a new multicast transmission strategy for mmWave communications.

Recent advances in the design of RF circuits in the $30-300$~GHz frequency range, along with studies on the corresponding propagation characteristics showed that using mmWaves for 5G cellular systems is actually feasible~\cite{Rappaport2013}. Thanks to their high transmission frequencies and correspondingly huge amount of available bandwidth, mmWave systems have the potential to solve the spectrum crunch. However, new design perspectives have to be introduced at the communication layers, since mmWave links are generally directional (in order to compensate for the strong path loss effect), and thus they present different characteristics with respect to traditional systems.

The physical properties of mmWaves present several benefits but also drawbacks. On the positive side, since the transmission beams are generally generated only toward particular directions, energy is not wasted in unwanted directions, and privacy and security concerns may be alleviated. Moreover, the interference to other nodes is reduced, allowing higher information capacities on the links.
Nevertheless, the main drawbacks of mmWave communications are the huge propagation loss due to the high transmission frequencies~\cite{Akdeniz2014} (according to Friis formula, the attenuation may easily increase by $30-40$~dB) and the blockage effect (because of the weak diffraction capabilities of waves, obstacles such as human being or furniture may heavily impair the communication link). Moreover, the directionality properties require more complex network discovery~\cite{Park2015,Giordani2016} and multicast algorithms~\cite{Park2013}.

Although many previous papers on mmWave communications focused on the optimization of data transmission using \emph{unicast} links~\cite{Niu2015}, in this paper we address \emph{multicast} communications~\cite{Park2013}.
These consist in transmitting the same data packets to a group of mobile terminal by performing the transmission only once, which consequently improves the bandwidth efficiency compared to unicast transmissions. Many different uses of multicast communications have been proposed in the past~\cite{Vella2013}, e.g., multimedia applications, distance learning, streaming of live events, VoIP sessions, etc., but also for control plane, synchronization, or reliability purposes. If correctly designed, also mmWave links can exploit multicasting; however, there are new aspects to take into account. For example, while in traditional networks beams are omnidirectional, with mmWave we only have directional beams, therefore we need to tune the beamwidth and the beam orientation in addition to the beam radius (in practice, multicast transmissions using directional links are a generalization of the traditional broadcast schemes)~\cite{Hou2007}.

Previous papers focused on multicast with directional links~\cite{Hou2007,Sundaresan2009}. However, these did not consider the possibility of tuning the beamwidth, which instead is a key factor in mmWaves. In~\cite{Park2013,Naribole2016}, the authors focused on multicasting for mmWave, but they did not consider the probability of losing packets during the transmissions, nor the effect of  retransmissions. Moreover, they did not optimize the number of packets to transmit over every beam. Instead, in our paper we implement a Hybrid Automatic Repeat reQuest (HARQ) mechanism, as a way to mitigate packet losses due to bad channel conditions.

\emph{\textbf{Our Contributions.}} We study a mmWave communication system in which a base station sends multicast packets to a group of users by tuning the parameters of the transmission beams over time. Lost packets are recovered by a HARQ mechanism, implemented with an incremental redundancy packet-level FEC code. The goal is i) to transmit the packets before a time deadline occurs and ii) to minimize the channel usage times. We solve the problem using a Markov Decision Process (MDP)~\cite{Bertsekas2005}, and, because of the super-exponential complexity of the optimal solution, we propose a simpler hierarchical optimization strategy. Numerical results show that multicast communications can significantly improve the system performance with respect to unicast only links.

The paper is organized as follows. Section~\ref{sec:sys_model} defines the system model. In Section~\ref{sec:opt_problem} we describe the multicast optimization problem and solve it optimally. A suboptimal solution is given in Section~\ref{sec:subopt_sol}. Section~\ref{sec:num_results} provides the numerical results and Section~\ref{sec:conclusions} concludes the paper.

\section{System Model}\label{sec:sys_model}

A Base Station (BS) transmits data to multiple users using a multicast wireless mmWave link. BS uses analog beamforming, so that it only irradiates a single beam to serve the users; with hybrid beamforming, the scheme proposed in this paper can be straightforwardly used for multiple separate beams, taking into account the total power constraint~\cite{Shokri-Ghadikolaei2015a}. Our goal is to tune the width of the transmission beam and its orientation in order to find the optimal trade-off between serving multiple users simultaneously and providing high data rates to them. Indeed, although a wider beam covers a larger spatial area and thus, potentially, may transmit data to many users in a short time, it also provides lower SNRs at the receiver (i.e., user) sides~\cite{Shokri-Ghadikolaei2015} and induces higher packet loss rates. Vice-versa, using a series of unicast beams provides higher SNRs but also requires longer transmission times.

The network is composed of $N$ end-users which demand to receive the same data packet from BS in downlink.\footnote{When several packets have to be transmitted, we apply the same transmission procedure for each of them. Thus, in this paper we only focus on the transmission of a single packet without loss of generality.} The base station modifies the position and the size of its transmission beam over time in order to satisfy the requests of all users. Since we consider multicast transmissions, a single beam may cover multiple users simultaneously.

When a packet is not received because of channel errors, it may be retransmitted. 
In particular, every data packet is encoded with a packet level FEC~\cite{Huitema1996} in smaller MAC packets, which are then transmitted independently. If a user correctly receives at least $m$ MAC packets, then decoding is possible (the value of $m$ depends on the code and on the size of the data packet). Formally, when BS transmits $x_i$ MAC packets, only $y_i \leq x_i$ are correctly received by User~$i$; if $y_i \geq m$, the data packet can be successfully decoded, and no other actions are required for that user. Otherwise, when $y_i < m$, decoding is not possible, and the base station may generate new additional redundancy packets and transmit them, until User~$i$ receives at least $m$ packets, or a time deadline is met.

Time is slotted, and slot $t$ denotes the normalized time interval $[t,t+1)$. In order to guarantee a bounded latency, every data packet admits at most $R_{\max}$ retransmission rounds before being declared ``failed'', i.e., if a user does not manage to decode the data packet within the first $R_{\max}+1$ slots, the time deadline is reached. In this situation, the system receives a penalty $\epsilon > 0$. The overall penalty, namely $\mathcal{E}$, is equal to $\epsilon$ times the number of users that reach the deadline.

When the system begins to operate at $t= 0$, every node receives a new data packet. The transmission of a single MAC packet to User~$i$ lasts $\tau(M_i)$ and depends on the selected modulation along with its channel code rate, $M_i$. The system strives to use shorter transmission durations, so that the base station is allowed to perform also other tasks, which are not explicitly defined here (e.g., transmission to other nodes, unicast transmissions, beam synchronization, localization, etc.). Thus, the longer the transmission times, the worse the system performance. On the other hand, shorter transmission times may induce higher packet loss rates, thus increasing the probability that the penalty $\epsilon$ is incurred. The goal of our optimization will be to find the trade-off between these two opposites.

\subsection{Beams}

For numerical tractability, we approximate the mmWave beam with a sectored antenna beam as in~\cite{Wildman2014,Bai2014}. When a receiver lies inside the sectored antenna beam, it receives a power given by Equation~\eqref{eq:P_received}; otherwise, no signal can be received.

When a beam with beamwidth $\psi$ is used, the received signal power of User~$i$ is, according to Friis formula,~
\begin{align}\label{eq:P_received}
    P_i = P_{\rm tx} \, h_i \, G_{\rm tx}(\psi) \, G_{\rm rc} \, PL_0 \, d_i^{-\alpha},
\end{align}

\noindent where $P_{\rm tx}$ is the transmission power of BS, assumed fixed in this paper and equal for all beams; $h_i$ is the random fading coefficient between BS and User~$i$ (independent over time and among users); $G_{\rm tx}(\psi)$ and $G_{\rm rc}$ are the transmitter and receiver antenna gains, respectively, in the direction of each other; $d_i$ is the distance between BS and User~$i$, $PL_0 = (\lambda/(4\pi))^2$ is the reference path loss at the distance of $1$~m ($\lambda$ is the wavelength of the signal), and $\alpha$ is the path loss exponent. We also notice that all these variables may be considered time dependent; although our model and its solution are general and would still be valid in this case, we keep them fixed for presentation simplicity.
The transmitter antenna gain depends on the beamwidth $\psi$ (the larger $\psi$, the lower $G_{\rm tx}(\psi)$) according to the following formula~\cite{Shokri-Ghadikolaei2015}:~
\begin{align}\label{eq:G_tx}
    G_{\rm tx}(\psi) = \frac{2\pi - (2\pi - \psi)z}{\psi},
\end{align}

\noindent where $0 \leq z \ll 1$ is the gain in the side lobe (e.g., $z = 0.05$). The previous expression holds only if the receiver lies in the coverage area of the beam. Note that, if $\psi = 2\pi$ (omnidirectional antenna), the transmitter gain $G_{\rm tx}(\psi)$ is minimum. The minimum beamwidth $\psi_0$, namely \emph{resolution}, is imposed by the number of antennas of the base station.

The SNR at user $i$ will be $P_i/(N_0\, W)$, where $N_0$ is the noise power spectral density and $W$ is the bandwidth. Depending on the SNR and on the modulation scheme, a user will experience a certain packet loss probability. In particular, let $\mathcal{M}$ be the set of allowed modulations along with their channel code rates (e.g., $\mathcal{M} = \{\mbox{BPSK with rate }1/2,\ \mbox{QPSK with rate }3/4\}$); in general, higher order modulations correspond to lower transmission durations but also to higher packet loss probabilities. Choosing the proper modulation scheme and code rate will be one of the objectives of the optimization problem.

\subsection{Markov Chain Formulation}\label{subsec:Markov}

The time horizon of our problem is imposed by the maximum number of retransmission rounds $R_{\rm max}$. At time $t = 0$, the base station transmits $x_i^0$ MAC packets to User~$i$; since only $y_i^0$ out of $x_i^0$ are received, at time $t = 1$, to correctly decode the data packet, User~$i$ needs to receive $r_i^1 \triangleq \max\{0,m-y_i^0\}$ packets. If $r_i^1 = 0$, the data packet can be decoded, otherwise additional transmissions are required. In a generic slot $t$, User~$i$ needs $r_i^t = \max\{0,r_i^{t-1}-y_i^{t-1}\}$ to decode. Thus, the system behavior in slot $t$ depends only on the previous slot state. Because of this, we can use a Markov Chain (MC) to model the retransmission states of the system.

When we consider all the $N$ users together, the state of the MC is $\mathbf{r}^t \triangleq \langle r_1^t,\ldots,r_N^t \rangle$, where $r_i^t,\ i \in \{1,\ldots,N\}$, is the number of packets that User~$i$ must receive to correctly decode the data packet in slot $t$. When a new data packet is generated at $t = 0$, then $r_i^0 = m, \ \forall i$. The system penalty is incurred when some user has not managed to decode the packet before the deadline, and is computed as~
\begin{align}\label{eq:Epsilon}
    \mathcal{E} \triangleq \epsilon \sum_{i = 1}^N \chi\big\{r_i^{R_{\max}+1} > 0\big\},
\end{align}

\noindent where $\chi\{\cdot\}$ is the indicator function.

\subsection{Actions}\label{subsec:actions}

Assume that the modulation along with the code rate are fixed. According to the position of the users, the penalty cost $\epsilon$, and the transmission parameters, BS can make the following choices in slot $t$:~
\begin{itemize}
    \item \textbf{Unicast Only.} Transmit $x_i^t \geq r_i^t$ unicast packets to User~$i$, for every $i$, using $N$ directional beams in a TDMA fashion. This will require $\sum_{i = 1}^N x_i^t\,\tau(M_i^t)$ seconds (we recall that $M_i^t$ is the modulation along with the code rate chosen for User~$i$ in slot $t$, and that the transmission duration $\tau(M_i^t)$ depends on $M_i^t$ only). Note that we may choose $x_i^t$ strictly greater than $r_i^t$ to take into account a non-zero probability of channel errors; in an error-free channel, we would always set $x_i^t = r_i^t$.
    \item \textbf{Broadcast.} Transmit $x_{1,\ldots,N}^t \geq \max\{r_1^t,\ldots,r_N^t\}$ broadcast packets to all users $\{1,\ldots,N\}$ using a large transmission beam. This will require $x_{1,\ldots,N}^t\,\tau(M_{1,\ldots,N}^t)$ seconds for the transmission. 
    In general, using a broadcast beam provides shorter transmission times, but also leads to worse SNR performance, therefore the packet loss rates increase and additional retransmissions may be necessary in the future (e.g., if the distance between users is wide, broadcast transmissions may be infeasible in mmWave).
    \item \textbf{Sequential Multicast.} Transmit a series of multicast beams in a TDMA fashion. Every beam covers only a portion of the users (eventually, even unicast beams may be employed). Note that in a single slot the same user may be covered by multiple beams (e.g., a unicast beam to increase the error resilience and a multicast beam to serve more users simultaneously and reduce the transmission times).
\end{itemize}

The goal of our optimization is to derive the optimal actions (i.e., the beam sequence) to perform in the sequential multicast case,\footnote{Clearly, the  unicast and broadcast schemes can be seen as particular cases of the sequential multicast one, therefore we do not discard pure unicast or broadcast solutions.} in every time slot. In addition, we also optimize the modulation and code rates for every transmission (intuitively, multicast packets may need more robust modulations, whereas unicast transmissions can use higher order modulation schemes).

\section{Optimization Problem and Solutions}\label{sec:opt_problem}

The goal of the system is twofold. First, we want to correctly send the data packet to all $N$ nodes in the first $R_{\rm max}+1$ slots (i.e., before the time deadline occurs). Second, we do not want to waste time resources; indeed, coding and transmitting a very large number of MAC packets may easily satisfy the deadline requirement, but would also incur long transmission times. Vice-versa, if very few packets were sent, the overall transmission duration would be very low but the deadline penalty may easily occur.

\subsection{Formal Optimization Problem}

The previous trade-off can be handled as an average multi-objective stochastic undiscounted finite-horizon optimization problem. The weight is given by the penalty $\epsilon$, which intrinsically determines whether we prefer to transmit few packets and save time or to risk not meeting the deadline. The problem is ``stochastic'' because of the unknown channel conditions, which may cause packet losses, and ``finite-horizon'' because we focus only on the first $R_{\rm max}+1$ slots. Moreover, the problem depends on the position of the users, therefore different configurations lead to different beam allocation.
Formally, the problem is defined as follows:
\begin{subequations}
    \label{eq:opt_problem}
    \begin{align}\label{eq:opt_prob_min}
        \begin{split}
            \min_{\substack{
                x_{\mathcal{N}}^t, \, M_{\mathcal{N}}^t \\
                \forall \mathcal{N} \subseteq \{1,\ldots,N\} \\
                t = 0,\ldots,R_{\rm max}
            }} &\mathbb{E}\Big[\sum_{t = 0}^{R_{\max}} \sum_{i=1}^N \sum_{\mathcal{N}\,:\, i \in \mathcal{N}} x_{\mathcal{N}}^t \, \tau(M_{\mathcal{N}}^t) \\
            &+ \epsilon \sum_{i = 1}^N \chi\big\{r_i^{R_{\max}+1} > 0\big\} \Big],
        \end{split}
    \end{align}
    \vspace{-\belowdisplayskip}
    \vspace{-\abovedisplayskip}
    \begin{alignat}{2}
        \shortintertext{s.t.:}
        & r_i^t = \begin{cases}
            m, \quad & \mbox{if } t = 0,\\
            \max\{0,r_i^{t-1} - \sum_{\mathcal{N}\,:\, i \in \mathcal{N}} y_{\mathcal{N}}^{t-1}\}, \quad & \mbox{if } t > 0,
        \end{cases} \label{eq:r_i_t} \\
        & \mathbb{P}(y_{\mathcal{N}}^t | x_{\mathcal{N}}^t, M_{\mathcal{N}}^t) = {{x_{\mathcal{N}}^t}\choose{y_{\mathcal{N}}^t}} \mathbb{P}_{\rm dec}(M_{\mathcal{N}}^t\big)^{y_{\mathcal{N}}^t} \overline{\mathbb{P}_{\rm dec}}(M_{\mathcal{N}}^t)^{x_{\mathcal{N}}^t-y_{\mathcal{N}}^t}, \label{eq:P_y_N_t}\\
        & x_{\mathcal{N}}^t \in \mathbb{N}, \quad\quad \forall \mathcal{N} \subseteq \{1,\ldots,N\},\ t = 1,\ldots,R_{\rm max}, \\
        & M_{\mathcal{N}}^t \in \mathcal{M}, \quad\, \forall \mathcal{N} \subseteq \{1,\ldots,N\},\ t = 1,\ldots,R_{\rm max}.
    \end{alignat}
\end{subequations}

\noindent The integer $x_{\mathcal{N}}^t$ represents the number of MAC packets transmitted in slot $t$ over a beam that covers all nodes in $\mathcal{N}$ (e.g., $x_1^t$ represents a unicast transmission to User~$1$, whereas $x_{1,\ldots,N}^t$ is a multicast transmission to all users). The variable $M_{\mathcal{N}}^t$ defines the modulation scheme along with its channel code for the beam that covers set $\mathcal{N}$ in slot $t$. Thus, the first sums in~\eqref{eq:opt_prob_min} represent the cost incurred for the transmissions. The higher the number of transmitted packets, the higher the transmitted durations and thus the cost in the objective function. Instead, the second term is the cost incurred for a missed deadline (defined in Equation~\eqref{eq:Epsilon}). The expectation is taken with respect to the channel conditions.

The set $\mathcal{N}$ contains the indices of the nodes. For example, it may be equal to $\{i\}, \, \forall i$, or to $\{i,j\}, \, \forall i, j>i$, \ldots, or to $\{1,\ldots,N\}$. Although it is not explicitly written in Problem~\eqref{eq:opt_problem}, we consider only ordered sets (e.g., $\{1,2\}$ is a valid $\mathcal{N}$, whereas $\{2,1\}$ is not) in order not to count the same beam multiple times. Note that the \emph{directionality} of the beam is already included in the definition of $\mathcal{N}$.

Equation~\eqref{eq:r_i_t} is the extension of what we presented in Subsection~\ref{subsec:Markov} to multiple beams. $r_i^t$ is the number of MAC packets that User~$i$ still has to transmit in slot $t$ to correctly decode the data packet. Correspondingly, \eqref{eq:P_y_N_t} is the probability of receiving $y_{\mathcal{N}}^t$ packets out of $x_{\mathcal{N}}^t$ transmitted, when a modulation $M_{\mathcal{N}}^t$ is used. The distribution of $y_{\mathcal{N}}^t$ is Binomial, with parameters $x_{\mathcal{N}}^t$ and $\mathbb{P}_{\rm dec}(M_{\mathcal{N}}^t)$ (probability of decoding a MAC packet). Note that we implicitly imposed independence among different packets, as the fading conditions change over time. The probability $\mathbb{P}_{\rm dec}(M_{\mathcal{N}}^t)$ depends on the link budget, the channel conditions, the modulation scheme, and the channel code. In particular, it is evaluated as $\mathbb{P}_{\rm dec}(M_{\mathcal{N}}^t) = \mathbb{E}[\mathbb{P}_{\rm dec}(M_{\mathcal{N}}^t,P_i)]$, where the expectation is taken with respect to the channel condition, $P_i$ is defined in~\eqref{eq:P_received}, $\mathbb{P}_{\rm dec}(M_{\mathcal{N}}^t,P_i)$ is the traditional packet decoding probability of a given modulation, and $\overline{\mathbb{P}_{\rm dec}}(M_{\mathcal{N}}^t) = 1 - \mathbb{P}_{\rm dec}(M_{\mathcal{N}}^t)$.

Finally, note that Problem~\eqref{eq:opt_problem} implicitly makes the conservative assumption that the beam directed to group $\mathcal{N}$ is not received by any other node. In practice, also other nodes may benefit from the data sent using this beam (e.g., if two or more users were close), therefore the performance of the system may slightly improve in practice.

In the next subsection, we describe how to solve the optimization problem optimally. Then, in Section~\ref{sec:subopt_sol} we will introduce a more faster solution which can be used in practice.

\subsection{Optimal Solution of Problem~\eqref{eq:opt_problem}} \label{subsec:opt_sol}

Intuitively, the optimal transmission policy that solves~\eqref{eq:opt_problem} tries to transmit few packets in the first retransmission rounds, so as not to increase the transmission durations (i.e., the first term in~\eqref{eq:opt_prob_min}), whereas it transmits more MAC packets in the last slots, if necessary, so as to avoid the penalty cost $\mathcal{E}$ (i.e., the second term in~\eqref{eq:opt_prob_min}).

\begin{figure}[t]
    \begin{center}
        \includegraphics[width=.8\columnwidth]{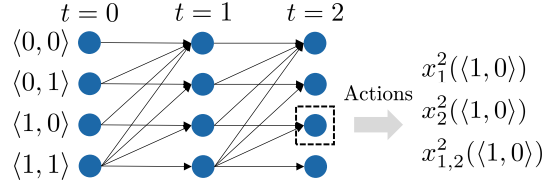}
        \caption{Markov Chain states when $N = 2$, $m = 1$, and $R_{\rm max} = 2$. For example, when $t = 2$ and in state $\mathbf{r} = \langle 1,0 \rangle$, we must define the number of packets to send to User~$1$, to User~$2$ and to Users~$1$ and $2$ simultaneously.}
        \label{fig:MC}
    \end{center}
\end{figure}

Since the system can be modeled with a Markov Chain, as described in Subsection~\ref{subsec:Markov}, we can solve Problem~\eqref{eq:opt_problem} as a Markov Decision Process (MDP) over a finite horizon. In particular, the optimization problem is stochastic, thus, to fully solve it, we need to specify an action to perform for every different state of the MC, namely $\mathbf{r} \triangleq \langle r_1,\ldots,r_N \rangle$ (note that we dropped the time index superscript because the set of MC states does not change over time). Therefore, we explicitly write $x_{\mathcal{N}}(\mathbf{r})$ and $M_{\mathcal{N}}(\mathbf{r})$ to indicate that these quantities are referred to state $\mathbf{r}$. In \figurename~\ref{fig:MC}, we show an example of the states of the MC. For every state, we need to optimize the number of packets to send over each beam.~

A common tool to solve MDPs over finite horizons is the Value Iteration Algorithm (VIA)~\cite{Bertsekas2005}, which consists in rewriting the objective function in a recursive form, and solving every recursive equation, namely Bellman equation, independently. Formally, the Bellman equation is expressed as follows for every $t = 0,\ldots,R_{\rm max}$,~
\begin{align} \label{eq:Bellman}
    J_t(\mathbf{r}) = \min_{\substack{x_{\mathcal{N}}^t(\mathbf{r}), \, M_{\mathcal{N}}^t(\mathbf{r}) \\
            \forall \mathcal{N} \subseteq \{1,\ldots,N\}
        }} \Big\{&\sum_{i=1}^N \sum_{\mathcal{N}\,:\, i \in \mathcal{N}} x_{\mathcal{N}}^t(\mathbf{r}) \, \tau(M_{\mathcal{N}}^t(\mathbf{r})) \nonumber\\
    & + \sum_{\mathbf{r}'} \mathbb{P}(\mathbf{r}' | \mathbf{r}) J_{t+1}(\mathbf{r}') \Big\}
\end{align}

\noindent and $J_{R_{\rm max}+1}(\mathbf{r}) \triangleq \epsilon \sum_{i=1}^N \chi\{r_i > 0\}$. It can be easily verified that the objective function~\eqref{eq:opt_prob_min} is equal to $J_0(\langle m,\ldots,m \rangle)$. In the Bellman equation, the first term constitutes the instantaneous cost obtained in slot $t$ when the actions $\{x_{\mathcal{N}}^t(\mathbf{r}),M_{\mathcal{N}}^t(\mathbf{r}),\, \forall \mathcal{N}\}$ are used, whereas the last term corresponds to the future costs. The term $\mathbb{P}(\mathbf{r}' | \mathbf{r})$ specifies the probability of going to state $\mathbf{r}'$ from the previous state $\mathbf{r}$, given the actions (which, for ease of notation, are implicitly embedded in $\mathbb{P}(\mathbf{r}' | \mathbf{r})$) and is derived according to~\eqref{eq:r_i_t} and~\eqref{eq:P_y_N_t}.

\subsection{The Price of Optimality}

The value iteration algorithm consists in iteratively solving~\eqref{eq:Bellman} in a backward fashion (i.e., starting from $t = R_{\rm max}$ and going to $t = 0$), and gives the optimal solution of the initial optimization problem. By doing so, we greatly reduced the computational complexity, since the problem can be studied separately for every slot. Nevertheless, there are two main pitfalls to face in order to minimize~\eqref{eq:Bellman}: first, the space of $\mathbf{r}$ scales exponentially with the number of users, thus the Bellman equation has to be solved many times; second, the number of sets $\mathcal{N}$ (i.e., the number of different beams) scales as $2^N-1$ (e.g., with three users we consider $\{1\}$, $\{2\}$, $\{3\}$, $\{1,2\}$, $\{2,3\}$, $\{1,3\}$, and $\{1,2,3\}$), thus there are many actions to optimize for every MC state.

In summary, finding the \emph{optimal} solution of the multicast problem we consider has super-exponential complexity in the number of users. Therefore, we need simpler techniques to simplify the computations.

\section{A Hierarchical Solution} \label{sec:subopt_sol}

We propose a suboptimal solution to the Bellman equation~\eqref{eq:Bellman} which is particularly suitable for the cases in which the users can be naturally divided in subgroups. In this section, we drop the time index $t$ for ease of notation.

\subsection{Tree Structure}

We consider a tree in which every isolated user constitutes a leaf, whereas groups of users are the internal nodes of the tree. Recursively, groups of nodes are put together to form upper layers of the tree, until the root node, which is composed of all users. Every node of the tree corresponds to a beam toward the corresponding group (e.g., leaves are unicast beams, whereas the root is a multicast beam that covers all the network).

We now introduce an optimization scheme that, using a top-down approach, analyzes and solves a series of easier optimization problems for every node of the tree. In particular, for each node we develop a series of simpler reduced MDPs, whose actions and states are tightly related to the children of the node.

\emph{\textbf{Notes on Optimality.}} The optimal solution of the network decomposed using the tree structure, namely the \emph{optimal tree policy}, imposes an upper bound to the cost of the initial optimal policy.\footnote{The tree structure intrinsically cuts some beams, since it is not possible to put together every combination of users.
Although the proposed approach inherently leads to suboptimal solutions, we may be able to significantly simplify the numerical evaluation and, if the tree structure is correctly designed, the hierarchical approach may have performance very close to the optimal one. Indeed, the structure of the tree is a design parameter and can be adapted to every configuration of users. More details about this design process will be part of our future work.} However, even finding the optimal tree policy is computationally demanding. Therefore, in the following, we propose a technique that, exploiting the hierarchical structure, provides an upper bound to the cost of the optimal tree solution and, consequently, to the initial optimal policy.

\subsection{Reduced Markov Chain Formulation}

We introduce a new \emph{reduced} Markov Chain for every node of the tree, which can be seen as a simplified version of what we described in Subsections~\ref{subsec:Markov} and~\ref{subsec:opt_sol}.

The state of node $\mathcal{N}$ in the reduced MC is $r_{\mathcal{N}}$. It is equal to $r_i$ if $\mathcal{N}$ is the leaf $\{i\}$ (as in Subsection~\ref{subsec:Markov}); otherwise, $r_{\mathcal{N}}$ is an \emph{aggregate} state defined as~
\begin{align}\label{eq:reduced_MC_state}
    &r_{\mathcal{N}} \triangleq \max\{ r_{\mathcal{C}_1},\ldots,r_{\mathcal{C}_p} \},
\end{align}

\noindent where $\{\mathcal{C}_1,\ldots,\mathcal{C}_p\}$ are the children of node $\mathcal{N}$.
For example, if we considered the node $\mathcal{N} = \{1,2,5,14,18\}$, with children $\mathcal{C}_1 = \{1,2\}$, $\mathcal{C}_2 = \{5,14\}$ and $\mathcal{C}_3 = \{18\}$, the aggregate MC state would be $r_{1,2,5,14,18} = \max\{r_{1,2},r_{5,14},r_{18}\}$ (instead of a five-dimensional one as in the initial MC). 

By doing so, the reduced MC always considers the worst case situation of the initial MC. In practice, \eqref{eq:reduced_MC_state} is a way to aggregate different states into a simpler one. Clearly, when these aggregate states are considered, we lose some information about the dynamics of the system.

\subsection{Aggregate Actions} \label{subsec:agg_actions}

There are many ways to define the actions of the aggregate states $r_{\mathcal{N}}$. In order to keep the optimization simple, we use, for every node of the tree, a number of actions equal to the number of children of the node under investigation plus two. 

One aggregate action, namely $a_{\mathcal{C}_\ell}$, is selected for every children, and two actions, $x_{\mathcal{N}}$ and $M_{\mathcal{N}}$, are reserved for defining the current beam (we recall that every node of the tree corresponds to a beam). Actions $x_{\mathcal{N}}$ and $M_{\mathcal{N}}$ are the number of MAC packets to send and the corresponding modulation to use, respectively, for the group of users $\mathcal{N}$ (similar to Subsection~\ref{subsec:actions}). Instead, when the aggregate action $a_{\mathcal{C}_\ell}$ is selected for child $\mathcal{C}_\ell$, then we impose an upper bound on the number of packets transmitted by the users in $\mathcal{C}_\ell$. For example, if we considered the node $\{1,2,5,14,18\}$, with children $\{1,2\}$, $\{5,14\}$ and $\{18\}$, the tuple of actions to decide would be $\{a_{1,2},a_{5,14},a_{18},x_{1,2,5,14,18},M_{1,2,5,14,18}\}$. Thus, for example, child $\{1,2\}$ imposes $a_{1} \leq a_{1,2}$, $a_{2} \leq a_{1,2}$, and $x_{1,2} \leq a_{1,2}$.

In practice, the bounds on the actions are propagated from the root to the leaves; by doing so, an action of an internal node indirectly influences the underlying subtree. Aggregate actions decide in which part of the network we should transmit more MAC packets, but without explicitly defining every beam (this will be done in a hierarchical fashion by lower layers), and thus are much easier to optimize.

\subsection{Transition Probabilities of the Reduced MC}\label{subsec:tx_prob_red_MC}

We need to define the transition probabilities among the aggregate states. Focus on a node $\mathcal{N}$ with children $\{\mathcal{C}_1,\ldots,\mathcal{C}_{p}\}$ and assume that the tuple of actions to perform $\{a_{\mathcal{C}_1},\ldots,a_{\mathcal{C}_p},x_{\mathcal{N}},M_{\mathcal{N}}\}$ is given.\footnote{The actions to perform are derived solving an MDP related to the reduced MC.} The aggregate transition probabilities are defined as follows~
\begin{subequations}
\label{eq:r_Nprime_GIV_r_N}
\begin{align}
    &\mathbb{P}(r_{\mathcal{N}}' | r_{\mathcal{N}},a_{\mathcal{C}_1},\ldots,a_{\mathcal{C}_p},x_{\mathcal{N}},M_{\mathcal{N}}) \\
    &= \sum_{r_{\mathcal{C}_1}} \cdots \sum_{r_{\mathcal{C}_p}} \sum_{r_{\mathcal{C}_1}'} \cdots \sum_{r_{\mathcal{C}_p}'} \chi\{\max\{r_{\mathcal{C}_1}', \ldots, r_{\mathcal{C}_p}'\} = r_{\mathcal{N}}'\} \\ 
    &\times \mathbb{P}(r_{\mathcal{C}_1}',\ldots,r_{\mathcal{C}_p}' | r_{\mathcal{C}_1}, \ldots, r_{\mathcal{C}_p}, a_{\mathcal{C}_1}, \ldots, a_{\mathcal{C}_p}, x_{\mathcal{N}},M_{\mathcal{N}}) \\
    &\times \chi\{\max\{r_{\mathcal{C}_1},\ldots,r_{\mathcal{C}_p}\} = r_{\mathcal{N}}\}.
\end{align}
\end{subequations}

\noindent In practice, the terms $r_{\mathcal{N}}$ and $r_{\mathcal{N}}'$ are decomposed according to their definitions in Equation~\eqref{eq:reduced_MC_state}. However, since $r_{\mathcal{N}}$ is an \emph{aggregate} state, we do not known the individual states of the children (i.e., $r_{\mathcal{C}_1},\ldots,r_{\mathcal{C}_p}$) but only the value of their maximum; in this case, we assume $r_{\mathcal{C}_1} = \ldots = r_{\mathcal{C}_p} = r_{\mathcal{N}}$, i.e., we consider the worst case scenario for computing the transition probabilities. Thus, Equation~\eqref{eq:r_Nprime_GIV_r_N} can be reduced to~
\begin{subequations}
\begin{align}
    &\mathbb{P}(r_{\mathcal{N}}' | r_{\mathcal{N}},a_{\mathcal{C}_1},\ldots,a_{\mathcal{C}_p},x_{\mathcal{N}},M_{\mathcal{N}}) \\
    &= \sum_{r_{\mathcal{C}_1}'} \cdots \sum_{r_{\mathcal{C}_p}'} \chi\{\max\{r_{\mathcal{C}_1}',\ldots,r_{\mathcal{C}_p}'\} = r_{\mathcal{N}}'\} \\ 
    &\times \mathbb{P}(r_{\mathcal{C}_1}',\ldots,r_{\mathcal{C}_p}' | r_{\mathcal{N}},\ldots,r_{\mathcal{N}},a_{\mathcal{C}_1},\ldots,a_{\mathcal{C}_p},x_{\mathcal{N}},M_{\mathcal{N}}). \label{eq:P_rCprime_rN}
\end{align}
\end{subequations}

\noindent The term~\eqref{eq:P_rCprime_rN} can be derived as~
\begin{subequations}
\begin{align}
    &\mathbb{P}(r_{\mathcal{C}_1}',\ldots,r_{\mathcal{C}_p}' | r_{\mathcal{N}}, \ldots, r_{\mathcal{N}}, a_{\mathcal{C}_1}, \ldots, a_{\mathcal{C}_p}, x_{\mathcal{N}},M_{\mathcal{N}}) \\
    &= \prod_{\ell = 1}^p \mathbb{P}(r_{\mathcal{C}_\ell}' | r_{\mathcal{N}}, a_{\mathcal{C}_\ell},x_{\mathcal{N}},M_{\mathcal{N}}) \\
    &= \prod_{\ell = 1}^p \sum_{u = 0}^{r_{\mathcal{C}_\ell}'} \mathbb{P}(r_{\mathcal{C}_\ell}' | u, x_{\mathcal{N}},M_{\mathcal{N}}) \mathbb{P}(u | r_{\mathcal{N}}, a_{\mathcal{C}_\ell}),
\end{align}
\end{subequations}

\noindent where in the last equality we used the total probability theorem. Term $\mathbb{P}(r_{\mathcal{C}_\ell}' | u, x_{\mathcal{N}})$ is the probability of going to the aggregate state $r_{\mathcal{C}_\ell}'$ starting from state $u$ given the actions $x_{\mathcal{N}}$ and $M_{\mathcal{N}}$, for child $\mathcal{C}_\ell$. This term can be evaluated using the definition of probability of the maximum along with Equation~\eqref{eq:P_y_N_t}. The other quantity, $\mathbb{P}(u | r_{\mathcal{N}}, a_{\mathcal{C}_\ell})$, is the probability of going to state $u$ starting from $r_{\mathcal{N}}$ using the aggregate action $a_{\mathcal{C}_\ell}$. This is derived from the solution of an MDP related to the reduced MC of child $\mathcal{C}_\ell$, given the constraint on the maximum number of MAC packets to transmit.

\subsection{Hierarchical Policy}

In the previous subsections, we have fully defined the MC related to the nodes of the tree. The hierarchical policy can be found by solving the MDP of the root of the tree corresponding to its reduced MC. This also implies to solve a series of MDPs for all the nodes of the tree, as explained in Subsection~\ref{subsec:tx_prob_red_MC}.

\section{Numerical Results} \label{sec:num_results}

In this section, we study how the system performance changes as a function of $m$ (number of MAC packets) and $R_{\rm max}$ (number of retransmission slots). The numerical results are derived in two stages. First, we follow the steps of Section~\ref{sec:subopt_sol} and perform a numerical evaluation to find the policy. Then, we use it to perform a Montecarlo simulation and assess the real performance of the system.

We adopt the following parameters. The link budget is modeled as in Equation~\eqref{eq:P_received}, where $\alpha = 3$, $G_{\rm rc} = 11.83$~dB (computed as in~\cite{Shokri-Ghadikolaei2015}), the central frequency is $28$~GHz, and $P_{\rm tx} = 1$~W. Also, we consider Nakagami fading with coefficient $4$. The bandwidth is $W=1$~GHz~\cite{Akoum2012} and the noise figure is $7.6$~dB. The beam parameters are $z = 0.05$ (power irradiated in the side lobes) and $\psi_0 = 11.25^{\circ}$ (resolution). The MAC packets are composed of $5$~KB plus $100$~bits of overhead. We consider two modulation schemes with Reed-Solomon codes $\mathcal{M} = \{\mbox{4-QAM with rate }239/255,\ \mbox{16-QAM with rate }223/255\}$. 

We initially consider a network composed of $8$ users with polar coordinates given in Table~\ref{tab:xy}. The tree of Section~\ref{sec:subopt_sol} is binary and the internal nodes are formed according to an increasing order of the indices of the users.

\begin{figure}[t]
    \begin{center}
        \includegraphics[width=1\columnwidth]{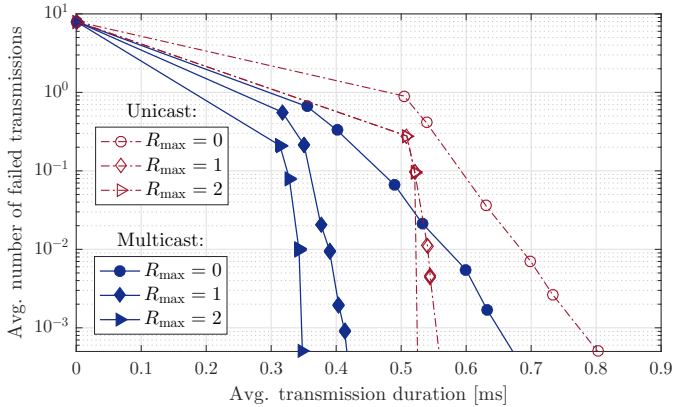}
        \caption{Average number of failed transmissions (i.e., users that have reached the deadline) as a function of the average transmission duration with $N = 8$ users, $m = 5$ MAC packets, and $0$, $1$, or $2$ retransmission slots. We remark that our problem (and in particular the action space) is \emph{discrete}, therefore it is not possible to operate for every combination of average transmission durations.}
        \label{fig:multi_vs_uni_change_R}
    \end{center}
\end{figure}

\begin{table}[h]
    \vspace{.4cm}
    \centering
    \caption{Position of the users.}
    \label{tab:xy}
    \begin{tabular}{ c|c|c|c|c|c|c|c|c }
        \toprule
        \textbf{User} & $1$ & $2$ & $3$ & $4$ & $5$ & $6$ & $7$ & $8$ \\
        \textbf{Radius} ($m$) & $100$ & $80$ & $50$ & $45$ & $30$ & $80$ & $100$ & $70$ \\
        \textbf{Angle} (degree) & $5$ & $25$ & $27$ & $35$ & $45$ & $65$ & $72$ & $86$ \\
        \bottomrule
    \end{tabular}
\end{table}

\figurename~\ref{fig:multi_vs_uni_change_R} shows the average number of users that reach the deadline as a function of the average transmission duration. The curves have been derived by changing the weight $\epsilon$ (the higher $\epsilon$, the lower the number of failures). In this example, we compare unicast and multicast policies. To derive the multicast scheme, we represented the $8$ users using the binary tree and used the algorithm of Section~\ref{sec:subopt_sol} to derive the solution. Instead, the unicast policies are derived optimally (in this case, the numerical complexity is very low, since every user can be analyzed independently of the others). It can be clearly noticed that, although the multicast scheme is suboptimal, it strongly outperforms the unicast one. Moreover, the optimal multicast approach might lead to even better performance than the hierarchical approach we used, thus the improvement of using multicast policies may be even higher.
In \figurename~\ref{fig:multi_vs_uni_change_R} we considered only $R_{\rm max} = 0,1,2$ retransmissions, because, for larger values, the performance almost saturates.

\begin{figure}[t]
\begin{center}
    \includegraphics[width=1\columnwidth]{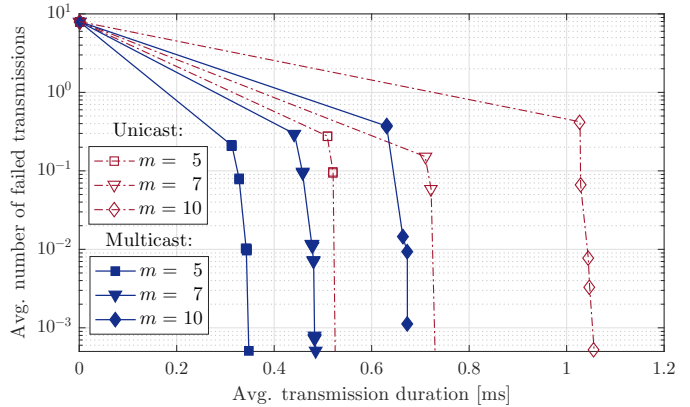}
    \caption{Average number of failed transmissions as a function of the average transmission duration with $N = 8$ users, $R_{\rm max} = 2$ retransmission slots, and $5$, $7$, or $10$ MAC packets.}
    \label{fig:multi_vs_uni_change_m}
\end{center}
\end{figure}

\figurename~\ref{fig:multi_vs_uni_change_m} is analogous to the previous one, but in this case we change the number of MAC packets required for decoding (e.g., higher values of $m$ correspond to larger data packets). When $m = 10$, many packets have to be transmitted, therefore it is more likely to incur in the time deadline penalty. 
Note that the gap between multicast and unicast policies is wider for higher values of $m$, which further justifies the use of multicast schemes for heavier data transmission applications (e.g., video streaming).

\begin{figure}[t]
\begin{center}
    \includegraphics[width=.7\columnwidth]{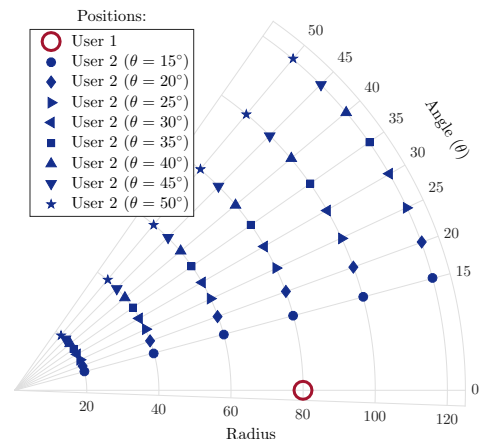}
    \caption{Positions of the users when $N = 2$ (User~$1$ is fixed, whereas $48$ different positions of User~$2$ are considered). The corresponding performance are shown in \figurename~\ref{fig:twoUsers}.}
    \label{fig:twoUsers_position}
\end{center}
\end{figure}

\begin{figure*}[t]
    \begin{center}
        \includegraphics[width=.33\linewidth]{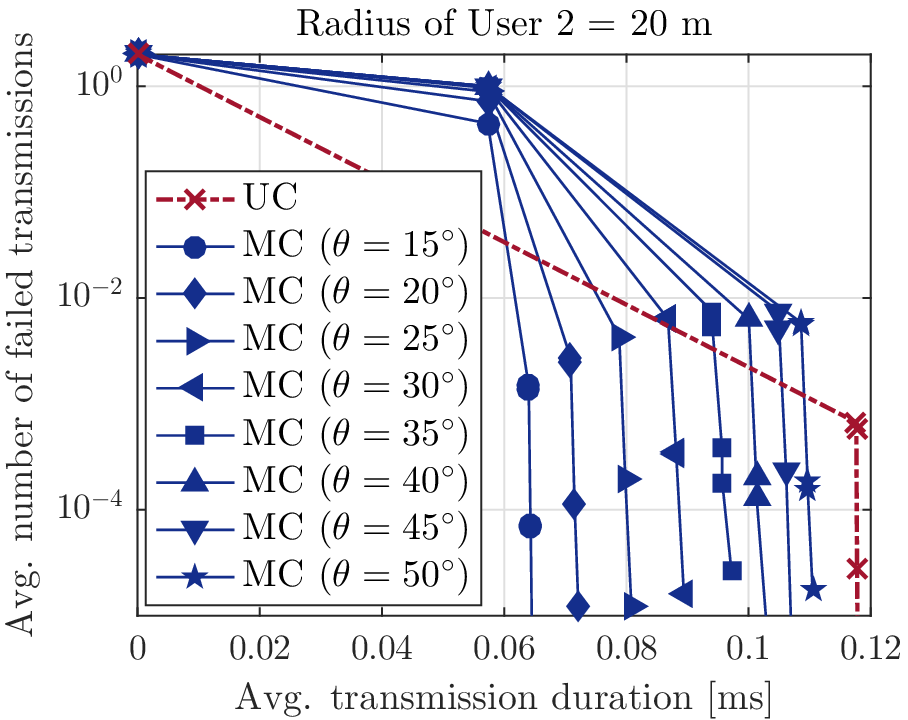}~
        \includegraphics[width=.33\linewidth]{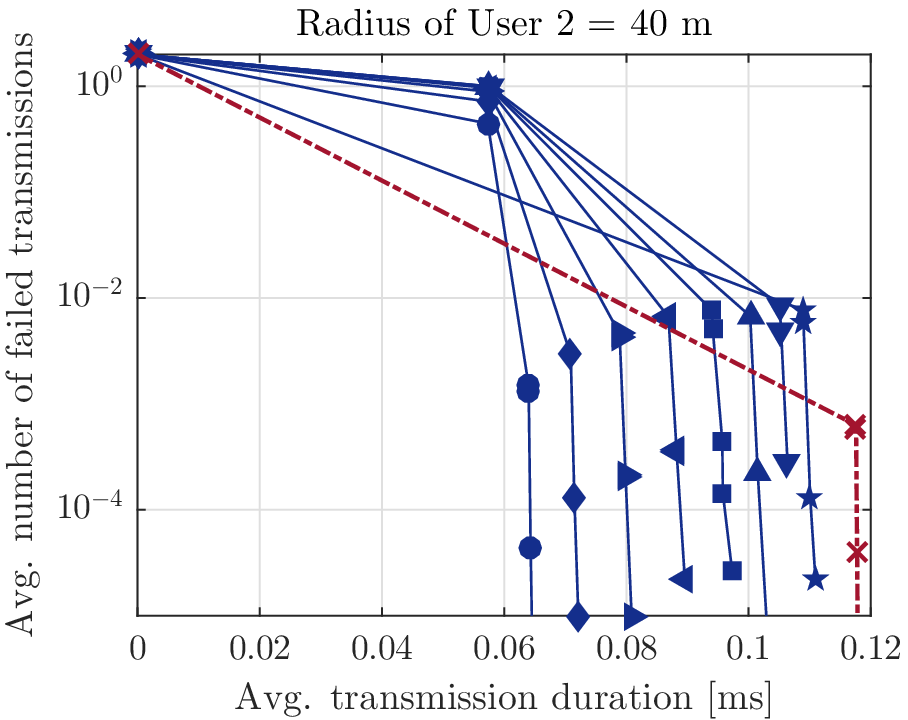}~
        \includegraphics[width=.33\linewidth]{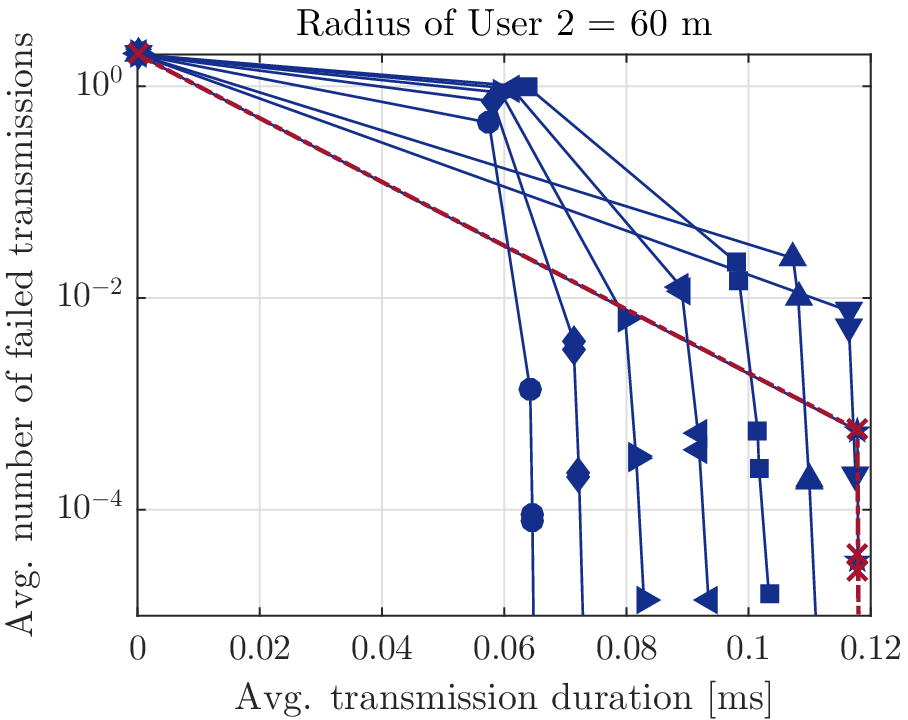}\\
        \vspace{.3cm}
        \includegraphics[width=.33\linewidth]{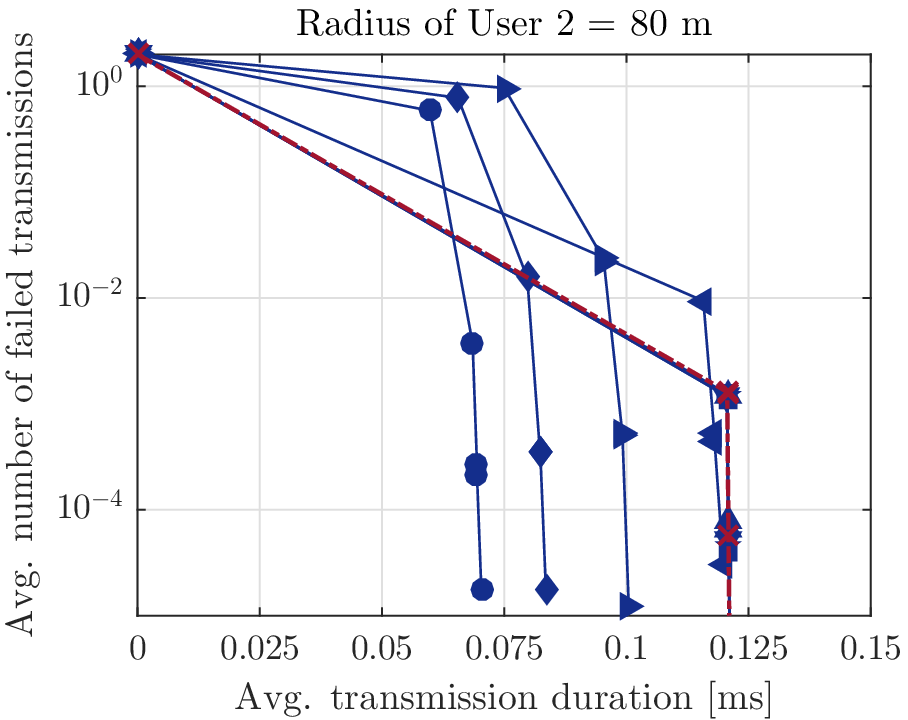}~
        \includegraphics[width=.33\linewidth]{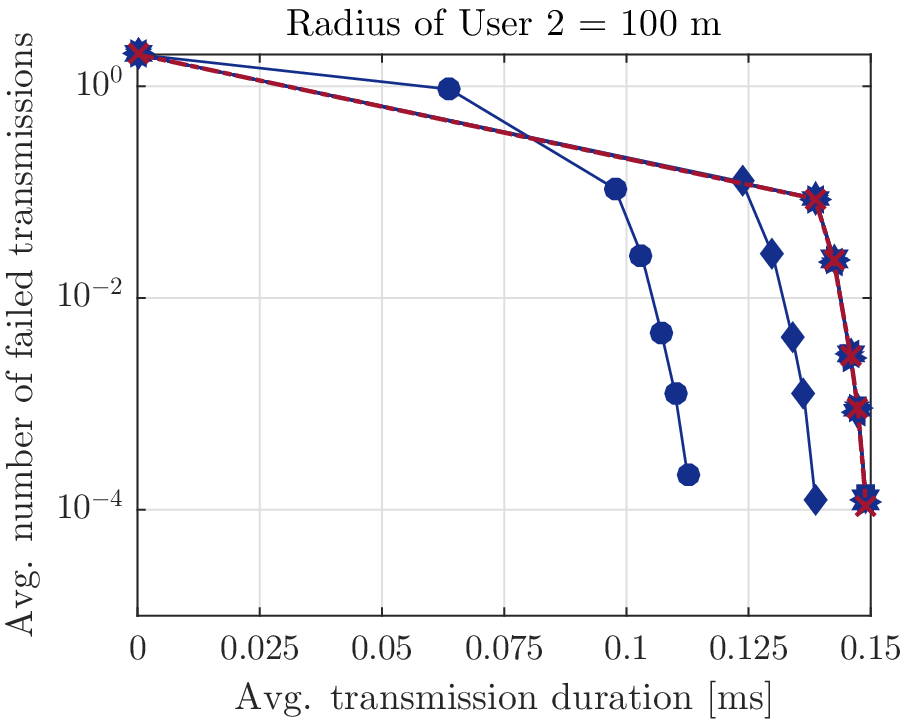}~
        \includegraphics[width=.33\linewidth]{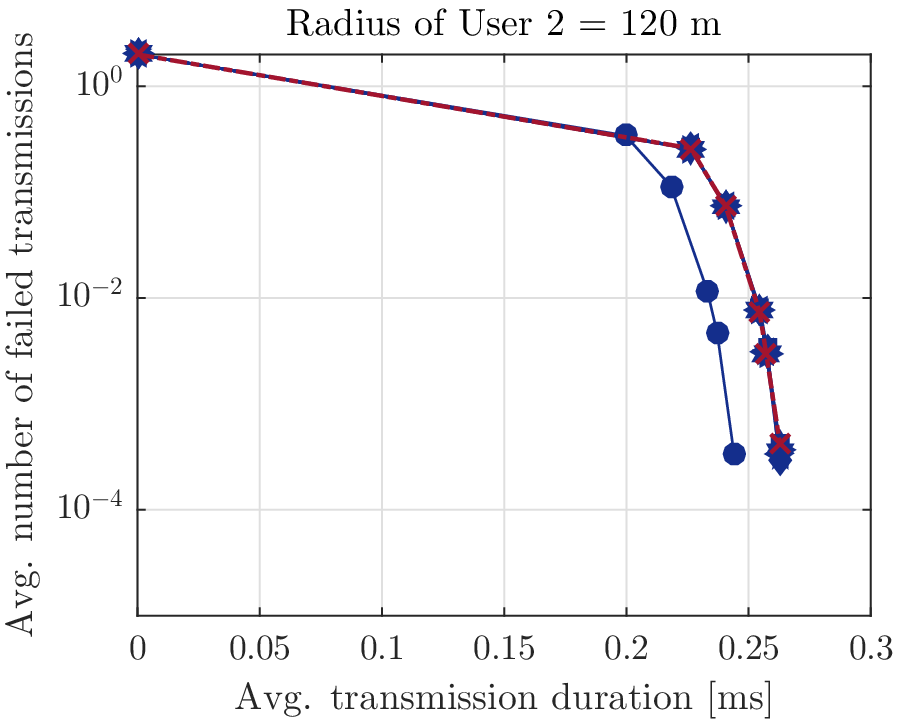}
        \caption{Average number of failed transmissions as a function of the average transmission duration with $N = 2$ users, $R_{\rm max} = 2$ retransmission slots, $m = 5$ MAC packets, and different positions of User~$2$ as shown in \figurename~\ref{fig:twoUsers_position}.}
        \label{fig:twoUsers}
    \end{center}
\end{figure*}

Finally, we focus on the simpler case of two users and consider different configurations as in \figurename~\ref{fig:twoUsers_position}. In particular, User~$1$ is placed at $(0^\circ,80~{\rm m})$ (polar coordinates), whereas User~$2$ is placed at many different positions. We do not consider very high (low) values of $\theta$ because unicast (multicast) schemes are almost always optimal in these cases. Moreover, the system is symmetric, therefore we could also have considered negative values of $\theta$ and the performance would have been the same. \figurename~\ref{fig:twoUsers} shows the corresponding performance of the system (UC and MC denote unicast and multicast transmissions, respectively). We remark that the unicast case (dashed curves) does not depend on the angle $\theta$, as a single beam covers only one user at a time in every case. As expected, the lower the value of $\theta$, the better the performance of the multicast scheme because the antenna gain for smaller beamwidths is higher (see Equation~\eqref{eq:G_tx}); in this case, the distance between multicast and unicast is significant. Also, we note that multicast performs better when the radius of User~$2$ is small; this is because the path loss is lower and it is possible to transmit MAC packets with a sufficiently low packet loss rate even using larger beams with smaller antenna gains. 

In summary, \figurename~\ref{fig:twoUsers} describes when it is worth using multicast in the two users case; when more users are considered, the benefits of multicast are even higher because a beam may cover more users. Moreover, according to \figurename s~\ref{fig:multi_vs_uni_change_R} and~\ref{fig:multi_vs_uni_change_m}, multicast becomes more important when $R_{\rm max}$ and $m$ increase, therefore, in practice, the improvement due to multicast may be expected to be very large in many cases of interest.

\section{Conclusions} \label{sec:conclusions}

We studied a multicast network composed of one base station that transmits packets with directional mmWave beams to multiple  users. We set up an optimization problem that balances the probability of failure with the average channel occupancy time. Since the optimal solution would require high numerical complexity (it requires to solve an MDP with a large state and action spaces), we reduced the complexity of the problem by introducing a hierarchical solution. We numerically evaluated the performance of the system with and without multicast beams, and noticed that, also in mmWave systems, multicast can significantly improve the performance of the system.

Part of our future work includes the comparison of the hierarchical approach with other suboptimal schemes, as well as with the optimal approach for some simple case. Moreover, additional investigations about the structure of the tree in the hierarchical solution may be performed.

\bibliography{bibliography}{}

\begin{thebibliography}{10}
\providecommand{\url}[1]{#1}
\csname url@samestyle\endcsname
\providecommand{\newblock}{\relax}
\providecommand{\bibinfo}[2]{#2}
\providecommand{\BIBentrySTDinterwordspacing}{\spaceskip=0pt\relax}
\providecommand{\BIBentryALTinterwordstretchfactor}{4}
\providecommand{\BIBentryALTinterwordspacing}{\spaceskip=\fontdimen2\font plus
\BIBentryALTinterwordstretchfactor\fontdimen3\font minus
  \fontdimen4\font\relax}
\providecommand{\BIBforeignlanguage}[2]{{%
\expandafter\ifx\csname l@#1\endcsname\relax
\typeout{** WARNING: IEEEtran.bst: No hyphenation pattern has been}%
\typeout{** loaded for the language `#1'. Using the pattern for}%
\typeout{** the default language instead.}%
\else
\language=\csname l@#1\endcsname
\fi
#2}}
\providecommand{\BIBdecl}{\relax}
\BIBdecl

\bibitem{Niu2015}
Y.~Niu, Y.~Li, D.~Jin, L.~Su, and A.~V. Vasilakos, ``A survey of millimeter
  wave {(mmWave)} communications for {5G:} opportunitios and challenges,''
  \emph{Wireless Networks}, vol.~21, no.~8, pp. 2657--–2676, Apr. 2015.

\bibitem{Vella2013}
J.-M. Vella and S.~Zammit, ``A survey of multicasting over wireless access
  networks,'' \emph{IEEE Commun. Surveys \& Tutorials}, vol.~15, no.~2, pp.
  718--753, Second Quarter 2013.

\bibitem{Rappaport2013}
T.~S. Rappaport, S.~Sun, R.~Mayzus, H.~Zhao, Y.~Azar, K.~Wang, G.~N. Wong,
  J.~K. Schulz, M.~Samimi, and F.~Gutierrez, ``Millimeter wave mobile
  communications for {5G} cellular: It will work!'' \emph{IEEE Access}, vol.~1,
  pp. 335--349, May 2013.

\bibitem{Akdeniz2014}
M.~R. Akdeniz, Y.~Liu, M.~K. Samimi, S.~Sun, S.~Rangan, T.~S. Rappaport, and
  E.~Erkip, ``Millimeter wave channel modeling and cellular capacity
  evaluation,'' \emph{IEEE J. Sel. Areas in Commun.}, vol.~32, no.~6, pp.
  1164--1179, June 2014.

\bibitem{Park2015}
H.~Park, Y.~Kim, T.~Song, and S.~Pack, ``Multiband directional neighbor
  discovery in self-organized {mmWave} ad hoc networks,'' \emph{IEEE Trans. on
  Vehicular Technology}, vol.~64, no.~3, pp. 1143--1155, Mar. 2015.

\bibitem{Giordani2016}
M.~Giordani, M.~Mezzavilla, and M.~Zorzi, ``Initial access in {5G mmWave}
  cellular networks,'' \emph{IEEE Communications Magazine}, vol.~54, no.~11,
  pp. 40--47, Nov. 2016.

\bibitem{Park2013}
H.~Park, S.~Park, T.~Song, and S.~Pack, ``An incremental multicast grouping
  scheme for {mmWave} networks with directional antennas,'' \emph{IEEE Commun.
  Letters}, vol.~17, no.~3, pp. 616--619, Jan. 2013.

\bibitem{Hou2007}
Y.~T. Hou, Y.~Shi, H.~D. Sherali, and J.~E. Wieselthier, ``Multicast
  communications in ad hoc networks using directional antennas: A
  lifetime-centric approach,'' \emph{IEEE Trans. on Vehicular Technology},
  vol.~56, no.~3, pp. 1333--1344, May 2007.

\bibitem{Sundaresan2009}
K.~Sundaresan, K.~Ramachandran, and S.~Rangarajan, ``Optimal beam scheduling
  for multicasting in wireless networks,'' in \emph{Proc. ACM 15th Annual Conf.
  on Mobile Computing and Networking (MobiCom)}, Sept. 2009, pp. 205--216.

\bibitem{Naribole2016}
S.~Naribole and E.~Knightly, ``Scalable multicast in highly-directional 60 {GHz
  WLANs},'' in \emph{Proc. IEEE 13th Annual Conf. on Sensing, Communication,
  and Networking (SECON)}, June 2016.

\bibitem{Bertsekas2005}
D.~Bertsekas, \emph{Dynamic programming and optimal control}.\hskip 1em plus
  0.5em minus 0.4em\relax Athena Scientific, Belmont, Massachusetts, 2005.

\bibitem{Shokri-Ghadikolaei2015a}
H.~Shokri-Ghadikolaei, C.~Fischione, G.~Fodor, P.~Popovski, and M.~Zorzi,
  ``Millimeter wave cellular networks: A {MAC} layer perspective,'' \emph{IEEE
  Trans. on Communications}, vol.~63, no.~10, pp. 3437--3458, July 2015.

\bibitem{Shokri-Ghadikolaei2015}
H.~Shokri-Ghadikolaei, L.~Gkatzikis, and C.~Fischione, ``Beam-searching and
  transmission scheduling in millimeter wave communications,'' in \emph{Proc.
  IEEE Conf. on Commun. (ICC)}, Sept. 2015, pp. 1292--1297.

\bibitem{Huitema1996}
C.~Huitema, ``The case for packet level {FEC},'' in \emph{Protocols for
  High-Speed Networks V}.\hskip 1em plus 0.5em minus 0.4em\relax Springer, Oct.
  1996, pp. 109--120.

\bibitem{Wildman2014}
J.~Wildman, P.~H.~J. Nardelli, M.~Latva-aho, and S.~Weber, ``On the joint
  impact of beamwidth and orientation error on throughput in directional
  wireless {Poisson} networks,'' \emph{IEEE Trans. Wireless Commun.}, vol.~13,
  no.~12, pp. 7072--7085, Dec. 2014.

\bibitem{Bai2014}
T.~Bai, A.~Alkhateeb, and R.~W. Heath, ``Coverage and capacity of
  millimeter-wave cellular networks,'' \emph{IEEE Communications Magazine},
  vol.~52, no.~9, pp. 70--77, Sept. 2014.

\bibitem{Akoum2012}
S.~Akoum, O.~El~Ayach, and R.~W. Heath, ``Coverage and capacity in {mmWave}
  cellular systems,'' in \emph{Proc. IEEE 46th Asilomar Conf. on Signals,
  Systems and Computers (ASILOMAR)}, Nov. 2012, pp. 688--692.

\end{thebibliography}
\bibliographystyle{IEEEtran}

\end{document}